\documentclass[twocolumn,prc,showpacs,preprintnumbers,
               unsortedaddress,amsmath,amssymb,floatfix]{revtex4}
\usepackage{graphicx}
\usepackage{dcolumn}
\usepackage{bm}
\usepackage{longtable}

\newcommand{\beq}{\begin{equation}}
\newcommand{\eeq}{\end{equation}}
\newcommand{\beqn}{\begin{eqnarray}}
\newcommand{\eeqn}{\end{eqnarray}}
\newcommand{\btab}{\begin{tabular}}
\newcommand{\etab}{\end{tabular}}

\newcommand{\ls}{\left[}
\newcommand{\rs}{\right]}

\usepackage{multirow,amsmath,booktabs}

\begin{document}

\title{Equation of State of Dense Matter from a density dependent relativistic mean field model}

\author{G.~Shen\footnote{e-mail:  gshen@indiana.edu}}
\affiliation{Nuclear Theory Center and Department of Physics,
Indiana University Bloomington, IN 47405}
\author{C.~J.~Horowitz\footnote{e-mail:
horowit@indiana.edu} } \affiliation{Nuclear Theory Center and
Department of Physics, Indiana University Bloomington, IN 47405}
\author{S.~Teige\footnote{e-mail:
steige@indiana.edu}}
\affiliation{University Information Technology Services, Indiana University, Bloomington, IN 47408}

\date{\today}
\begin{abstract}
We calculate the equation of state (EOS) of dense matter, using a relativistic mean field (RMF) model with a density dependent coupling that is a slightly modified form of the original NL3 interaction.  For nonuniform nuclear matter we approximate the unit lattice as a spherical Wigner-Seitz cell, wherein the meson mean fields and nucleon Dirac wave functions are solved fully self-consistently. We also calculate uniform nuclear matter for a wide range of temperatures, densities, and proton fractions, and match them to non-uniform matter as the density decreases. The calculations took over 6,000 CPU days in Indiana University's supercomputer clusters. We tabulate the resulting EOS at over 107,000 grid points in the proton fraction range $Y_P$ = 0 to 0.56.  For the temperature range $T$ = 0.16 to 15.8 MeV we cover the density range $n_B$ = 10$^{-4}$ to 1.6 fm$^{-3}$; and for the higher temperature range $T$ = 15.8 to 80 MeV we cover the larger density range $n_B$ = 10$^{-8}$ to 1.6 fm$^{-3}$. In the future we plan to study low density, low temperature (T$<$15.8 MeV), nuclear matter using a Virial expansion, and we will match the low density and high density results to generate a complete EOS table for use in astrophysical simulations of supernova and neutron star mergers.
\end{abstract}


\maketitle

\section{Introduction}

The equation of state (EOS) for hot, dense matter in massive stars
relates energy and pressure to temperature, density, and
composition.  It has been a long-standing problem to understand the
EOS at both subnuclear and supranuclear density, to which great
efforts have been devoted, from laboratory heavy ion collision
experiments \cite{heavyions}, computer simulations of supernova
\cite{simulation1,simulation2}, and theoretical many-body
calculations \cite{theory}.  The EOS of hot dense matter in
supernovae (SN) and neutron star (NS) mergers encompass multi-scale
physics.  Temperature can vary from 0 to as high as 100
MeV, exciting nuclei, nucleon and possibly pion and other degrees of freedom.  The density can vary from $\approx 10^{4}$ to 10$^{15}$ g$\cdot$ cm$^{-3}$, where matter can be in gas, liquid or solid phases.  The proton fraction can vary from 0 to 0.6, from extremely neutron rich matter to proton rich matter.  These very large parameter ranges make construction of a full EOS table difficult. It is necessary to employ different approximations for different parameter ranges. As a result, there exist only two realistic EOS tables that are in widespread use for astrophysical simulations, the Lattimer-Swesty (L-S) equation of state \cite{LS}, that uses a compressible liquid drop model with a Skyrme force, and the H. Shen, Toki, Oyamatsu and Sumiyoshi (S-S) equation of state \cite{Shen98a,Shen98}, that uses the Thomas Fermi and variational approximations with a relativistic mean field (RMF) model.  We plan to generate a complete equation of state, employing  relativistic mean field calculations for matter at intermediate and high density as described in this paper.  In the future we plan to use the Virial expansion of a nonideal gas to describe matter at low density.   These two parts will be matched together and we will generate a thermodynamically consistent EOS over the full range of parameters.  Finally we will generate additional EOSs from RMF models with different high density symmetry energies.  This will allow one to correlate features of astrophysical simulations with properties of the symmetry energy assumed for the EOS.


There are still large uncertainties in the EOS at supranuclear
densities.  The density dependence of the symmetry energy $dS/dn_B$ is poorly known and strongly influences the stiffness of the EOS.  It can be constrained from measurements of NS radii and masses \cite{Lattimer04}, precision determination of the neutron rms radius in $^{208}$Pb \cite{prex}, and also heavy ion collision experiments
\cite{heavyions}. A stiff EOS (high pressure) at high density gives larger NS
radii, while a stiff EOS at normal and low density favors a larger
neutron radius in $^{208}$Pb \cite{HP01}. The elliptic and
transverse flow observables in heavy ion collisions are sensitive
to the isospin dependence of mean fields and to the EOS at densities up to a few times
nuclear saturation density.  Many nuclear many-body
models fall into two categories, the non-relativistic Skyrme models
(See for example, Ref.~\cite{Talmi} for a review) and relativistic
mean field models \cite{RMFreviews}. The parameters in these
models are usually fitted to nuclear properties at normal nuclear
densities, afterwards they are extrapolated to study supranuclear
matter.  The L-S EOS uses a Skyrme model featuring a relatively soft EOS
and the S-S EOS uses the RMF interaction TM1 that features a stiffer
EOS.  Since the symmetry energy is not well constrained, it is
important to explore the effects of different symmetry energies on the EOS and SN
simulations.

In this paper, we use a RMF model for non-uniform matter at
intermediate density and uniform matter at high density.  Low density pure neutron matter is analogous to a unitary gas \cite{HS05b}, where the neutron-neutron scattering length is
much larger than both the effective range and the average inter-particle spacing. To better describe neutron-rich matter at low density, we use a density dependent scalar
meson-nucleon coupling. At high density, the model reduces to the normal RMF parameter set NL3.  The unit lattice of non-uniform nuclear matter is conveniently
approximated by a spherical Wigner-Seitz (W-S) cell. The meson mean
fields and nucleon Dirac wave functions inside the Wigner-Seitz cell are solved
fully self-consistently.  This is unlike the S-S EOS that used Thomas-Fermi and variational 
approximations and the L-S EOS that used a simple liquid drop model. The
size of the W-S cell is found by minimization of the free energy per
nucleon.  The W-S approximation provides a framework to incorporate the best known microscopic nuclear physics \cite{Negele73}. The nuclear shell structure effects
are included automatically and it is already possible for
some effects of complex nuclear pasta states to be included in spherical calculations in the form of shell states \cite{HS08}.  Full three-dimensional W-S calculations in principle could
incorporate various pasta shapes \cite{Gogelein07,Newton09}, which
would make the transition to uniform matter more smooth. However
this will demand much larger computational resources. In this work
we use the spherical W-S approximation.

Our relativistic mean field calculations can accurately describe the radial shape of large neutron rich nuclei including the expected neutron rich skin.  In contrast the original L-S EOS is based on a very simple liquid drop model of nuclear structure that may incorrectly describe the neutron skin.  Alternatively the S-S EOS is based on a Thomas Fermi approximation that neglects shell effects.  These are included in our Hartree calculations.  Furthermore, the variational forms for the densities assumed by S-S may be a poor approximation for large proton numbers where the Coulomb repulsion is large.  Instead our exact solutions of the radial mean field equations allow richer density distributions including shell states with central depressions \cite{HS08}.  These differences in densities may be important for neutrino interactions in Supernovae.  Finally our calculations correctly reproduce the Unitary gas limit for a low density neutron gas, see below, while both the L-S and S-S EOS reduce incorrectly to the energy of free neutrons.

One can demand that any EOS be consistent with, possibly model dependent interpretations of, observations of neutron stars.  For example, Klahn et. al. propose a series of tests that an EOS should satisfy to be consistent with observations \cite{klahn}.  They demand that any reliable nuclear EOS be able to reproduce the recently reported high pulsar mass of $2.1\pm$ 0.2 $M_\odot$ for PSR J0751+1807 \cite{nice}.  However, this observation may have been retracted \cite{notnice}.  Furthermore, Klahn et al. require the EOS to reproduce a large binding energy for Pulsar B in J0737-3039.  However, this conclusion could be sensitive to assumptions about the system such as the amount of mass loss.  Klahn et. al. go on to demand that the EOS not allow direct URCA cooling of neutron stars of mass 1 to 1.5 $M_\odot$.  We consider a more conservative approach.  While many stars cool slowly, observations do suggest that at least some stars have enhanced cooling.  Unfortunately observations do not directly constrain the mass that may separate enhanced from normal cooling.  Indeed, there is little direct observational evidence that more massive stars cool more quickly, although this is a theoretical prejudice.   

One can also use laboratory data to constrain the EOS.  The neutron skin thickness of a heavy nucleus constrains the density dependence of the symmetry energy.  Furthermore, there are many measurements of the skin thickness with a variety of strongly interacting probes.  However, there may be important model dependence from strong interaction uncertainties.  For example $(^3He,t)$ measurements of spin dipole strength have been used to extract neutron skin thicknesses in Sn isotopes \cite{spindipole}.  For these measurements, the spin dipole strength was assumed to be proportional to the measured cross section, and the proportionality constant was arbitrarily fixed in order to reproduce the skin thickness of $^{120}$Sn as predicted by an old Hartree-Fock calculation \cite{oldHF}.  Presumably, if a different skin thickness in $^{120}$Sn is fit, one would also get different answers for the skin thickness in other isotopes.   

This situation may soon change.  The Lead Radius Experiment (PREX) at Jefferson Laboratory is using parity violating electron scattering to measure the neutron skin thickness in $^{208}$Pb \cite{prex}.  Parity violation is a sensitive probe of neutrons because the weak charge of a neutron is much larger than that of a proton.  Furthermore, this electro-weak reaction may have much smaller strong interaction uncertainties.  Data taking for PREX should be completed by June 2010.

Instead of trying to determine, ahead of time, the best EOS to satisfy existing observational constraints, we adopt what we hope will be a more robust approach.  We are calculating a number of EOSs based on different effective interactions.  In this paper we present first results for the NL3 interaction with a symmetry energy that is large at high densities.  In later work we will present EOSs with softer high density symmetry energies.  These different EOS will allow one to correlate features of astrophysical simulations with properties of the EOS.  Then one can draw conclusions based on combined information from laboratory experiments and astronomical observations.    

In this paper we focus on nucleon degrees of freedom.  Hyperons could play a role at high densities, see for example \cite{shen2}.  However, the contribution of hyperons could depend on uncertain hyperon interactions.  In addition, there could be pion or kaon condensates or a variety of quark matter phases.  See for example the review by Page and Reddy \cite{theory}.  Chiral symmetry restoration and the softening of pionic or kaonic modes could be important.  Finally, thermal pions and pion interactions should be very important at high temperatures.   All of these effects may increase the uncertainties in the EOS.

We tabulate the equation of state for intermediate and high density nuclear matter over the range of temperatures $T$, densities $n_B$, and proton fractions $Y_P$ given in Table  \ref{tab:phasespace} and described in Sec. \ref{methodology}.  We calculate the free energy of nonuniform matter for 17021 points, and the free energy of uniform matter for 90478 points in $T$, $n_B$, and $Y_P$ space.  This took 6000 CPU days on Indiana University's supercomputer clusters.

The paper is organized as follows: in Section \ref{formalism}
the density dependent RMF model is explained in detail.  In Section \ref{numerics} we
describe the RMF parameters that we use including a density dependent coupling.  We describe the computational methodology for our large computer runs in Section \ref{methodology}.  Section \ref{result} shows results for RMF calculations, including the free energy and the nucleon density distributions in the non-uniform W-S cells.  Finally, Section \ref{summary} presents a summary of our results and gives an outlook for future work.

\begin{table}[h]
\centering \caption{Range of temperatures $T$, densities $n_B$, and proton
fractions $Y_P$ in the EOS table.} \label{tab:phasespace}\btab{cccc}
\hline \hline
Parameter & Low T & High T & Total \# \\
 \hline
log$_{10}$(T) [MeV] & -0.8 to 1.2 & 1.2 to 1.9 & 32 \\

log$_{10}$($n_B$) [fm$^{-3}$] &-4.0 to 0.2 & -8.0 to 0.2 & 43, 83 \\

$Y_P$  & 0,0.05 to 0.56 & \ 0,0.05 to 0.56  & 53 \\

\hline \etab
\end{table}

\section{\label{formalism}Formalism}
We now describe the mean field formalism that we use for non-uniform matter in Section \ref{subsec.nonuniform} and for uniform matter in Section \ref{subsec.uniform}.  

\subsection{Non-uniform nuclear matter in Wigner-Seitz
approximation}
\label{subsec.nonuniform}

The formalism for relativistic mean field theory has been reviewed
in previous papers, see eg \cite{RMFreviews}. To better describe
neutron rich matter at low density we introduce a density dependent coupling
between the scalar meson and the nucleon as described in Section \ref{numerics}.  We note that many previous studies of density dependent RMF models mainly focused on better descriptions of nuclear matter at supranuclear density (see for example,
\cite{Typel99,Ban04}).  In this section we focus on low density neutron
rich matter.

The basic ansatz of the RMF theory is a Lagrangian density where
nucleons interact via the exchange of sigma- ($\sigma$), omega-
($\omega_\mu$), and rho- ($\rho_\mu$) mesons, and also photons
($A_\mu$). \beqn\label{lagrangian}
    {\cal L}&=&\overline{\psi} [i{\gamma^\mu}
              {\partial_\mu}-m-{\Gamma_\sigma}\sigma - g_\omega
              \gamma^\mu\omega_\mu \nonumber\\
              &&-\
              g_\rho \gamma^\mu \vec{{\bf
              \tau}}\cdot \vec{\mbox{\boldmath$\rho$}}_\mu - e\gamma^\mu\frac
              {1+\tau_3}{2} A_\mu ]\psi\nonumber\\
            &&+\
            \frac{1}{2}\partial^\mu\sigma\partial_\mu\sigma-\frac{1}{2}m_\sigma^2\sigma^2-
              \frac{1}{3}g_2\sigma^3\ -\ \frac{1}{4}g_3\sigma^4\nonumber\\
            &&-\
              \frac{1}{4}\omega^{\mu\nu}\omega_{\mu\nu}+\frac{1}{2}m_\omega^2\omega^\mu\omega_\mu+
              \frac{1}{4}c_3\left(\omega^\mu\omega_\mu\right)^2\nonumber\\
            &&-\ \frac{1}{4}\vec{\mbox{\boldmath$\rho$}}^{\mu\nu}
              \cdot\vec{\mbox{\boldmath$\rho$}}_{\mu\nu}+
              \frac{1}{2}m_\rho^2\vec{\mbox{\boldmath$\rho$}}^\mu
              \cdot\vec{\mbox{\boldmath$\rho$}}_\mu -\
              \frac{1}{4}A^{\mu\nu}A_{\mu\nu}
\eeqn We note that $\Gamma_\sigma = \Gamma_\sigma(n)$
($n\equiv\sqrt{j_\mu j^\mu}$ and $j_\mu$ is nucleonic current) is
the density dependent coupling between the sigma meson and the nucleon.
Here the field tensors of the vector mesons and the
electromagnetic field take the following forms:
\beqn\label{tensor}
   \omega^{\mu\nu}&=&\partial^\mu\omega^\nu-\partial^\nu\omega^\mu,\nonumber\\
                     A^{\mu\nu}&=&\partial^\mu A^\nu -\partial^\nu A^\mu,\nonumber\\
                     \vec{\mbox{\boldmath$\rho$}}^{\mu\nu}&=&\partial^\mu\vec{\mbox{\boldmath$\rho$}}^\nu-
                     \partial^\nu\vec{\mbox{\boldmath$\rho$}}^\mu-g_\rho\vec{\mbox{\boldmath$\rho$}}^\mu\times
                     \vec{\mbox{\boldmath$\rho$}}^\nu\,.
\eeqn

In charge neutral nuclear matter composed of neutrons, $n$,
protons, $p$, and electrons, $e$, there are equal numbers of
electrons and protons.  Electrons can be treated as a uniform
Fermi gas at high densities {\footnote{It needs electron density
$> 10^6$ g/cm$^3$, which is easily surmounted in the regime of
mean field results.}}. They contribute to the Coulomb energy of
the $npe$ matter and serve as one source of the Coulomb potential.

The variational principle leads to the following equations of
motion 
\beqn\label{nucleon-motion}
     [\mathbf{\sl{\alpha}}\cdot\mathbf{\sl{p}}\ +\ V(\mathbf{r})\ +\
     \sl{\beta} (m\ +\ S(\mathbf{r}))] \psi_i\ =\ \varepsilon_i\psi_i
\eeqn for the nucleon spinors, with vector and scalar potentials
\beqn \label{Dirac}
     \begin{array}
        {l} V(\mathbf{r}) =\ \beta \{g_{\omega}\rlap{/}{\omega}_{\mu}
        + {g_{\rho}} \vec{\tau}\cdot
        \vec{\rlap{/}\rho_{\mu}} + e \frac{ (1 + \tau_3)}{2}\ \rlap{/}A_{\mu} + \Sigma^R \}, \\
        S(\mathbf{r}) =\ \Gamma_{\sigma}\sigma, \
     \end{array}
\eeqn 
where \beq \Sigma^R = \frac {\gamma^\mu j_\mu} {n}
\frac{\partial\Gamma_\sigma}{\partial n} \rho_s \sigma, \eeq is
the rearrangement term due to the density dependent coupling between the 
sigma meson and the nucleon, and $\rho_s$ is the scalar density of
nucleons to be defined below.

The equations of motion for the mesons and photons are,
\beqn\label{K-G}
    \begin{array}{l}
       (m_{\sigma}^2\ -\ \nabla^2)\
       \sigma =\ -\Gamma_\sigma \rho_s -\ g_2 \sigma^2\ -\ g_3\sigma^3, \\
       (m_\omega^2\ -\ \nabla^2)\omega^\mu =\ g_\omega j^\mu -c_3\omega^\mu(\omega^\nu\omega_\nu), \\
       (m_{\rho}^2\ -\ \nabla^2) \vec{\rho}^{\mu} =\ {g_{\rho}}
       \vec{j}^\mu, \\
       \label{speqa}\hspace{2.5em} -\ {\nabla}^2A^\mu =\
       e(j^\mu_{p}-j^\mu_e),
    \end{array}
\eeqn 
where the electrons are included as a source of Coulomb
potential.  The nucleon spinors provide the relevant source terms:
\beqn\label{source}
    \begin{array}{l}
      \rho_s =\ \sum_i\overline{\psi}_i\psi_i n_i, \\
      j^\mu =\ \sum_i\overline{\psi}_i \gamma^\mu \psi_i n_i,\\
      \vec{j}^\mu =\ \sum_i \overline{\psi}_i \gamma^\mu
      \vec{\tau} \psi_i n_i,\\
       j^\mu_{p} =\ \sum_i\overline{\psi}_i\gamma^\mu\frac
      {1+\tau_3}{2}\psi_i n_i.
\label{speqb}
    \end{array}
\nonumber\\
\eeqn

At finite temperature, Fermi-Dirac statistics imply the
occupations, $n_i$, of protons and neutrons are: \beq\label{occu}
    n_i\ =\ \frac 1 {e^{\beta(\varepsilon_i-\mu)}+1},
\eeq where $\mu$ is the chemical potential for neutron (proton).
In the calculation, we include all levels with $g_i\cdot
n_i>10^{-2}$, where $g_i$ is the degeneracy of the level.

Since the systems under consideration have temperatures of, at
most, tens of MeV, we neglect the contribution of negative energy
states, {\it i.e.}, the so-called no sea approximation. In a
spherical nucleus, there are no currents in the nucleus and the
spatial vector components of $\omega_\mu$, $\vec{\rho}_\mu$ and
$A_\mu$ vanish.  One is left with the timelike components,
$\omega_0$, $\vec{\rho}_0$ and $A_0$.  Charge conservation
guarantees that only the 3-component of the isovector $\rho_{0,3}$
field survives. The above non-linear equations are solved by iteration
within the context of the mean field approximation whereby the
meson field operators are replaced by their expectation values.

The spherical Wigner-Seitz approximation is used to describe
non-uniform matter.  One W-S cell has one nucleus. In this
approximation it is important to include lattice Coulomb
corrections between neighboring W-S cells. The detailed treatments
we use have been discussed in a previous paper \cite{HS08} and we will
not repeat them here.

We solve for the meson mean fields and the nucleon Dirac wave functions self-consistently inside a W-S cell of radius $R_c$, for a given baryon density
$n_B$, proton fraction $Y_P$ and temperature $T$. In our RMF model nucleons (proton and neutron) are the only baryons. The nucleon
number inside the W-S cell is $A = 4\pi R_c^3 n_B/3$
and the proton number is $Z = Y_P A$. The internal energy of a W-S cell,
including the approximate lattice Coulomb energy correction, is,
\begin{widetext}

\beqn\label{energy} E_b &=& E_{nucleon} + E_\sigma + E_\rho +
E_\omega + E_{Coul} - mA, \cr &=& \sum_i \epsilon_i n_i -\ \int
d^3r j_0(r)\frac{\partial\Gamma_\sigma}{\partial
j_0}\rho_s(r)\sigma(r) -\ \frac 1 2 \int d^3r \{\Gamma_\sigma
\sigma \rho_s(r)\ +\ \frac 1 3 g_2\sigma^3\ +\ \frac 1 2
g_3\sigma^4\},\ \cr && -\ \frac 1 2 \int d^3rg_\rho\rho_{0,3}
j_{0,3}(r)\ -\ \frac 1 2 \int d^3r\{g_\omega \omega_0\ j_0(r) -\
\frac 1 2 c_3\omega_0^4\} -\ \frac e 2 \int (\rho_p+\rho_e) A_0(r)
d^3r \ +\ dw- mA, \eeqn

\end{widetext}
where $dw = 0.0065620 Z^2/a$ is the approximate Coulomb correction for a bcc lattice\cite{Oyamatsu}, and $a^3 = V_{W-S}$ is the volume of W-S cell.

The nucleon contribution to the entropy is given by the usual
formula, \beq\label{entropy} S_b = -k_B \sum_i g_i \left[ n_i
\mathrm{ln} (n_i)\ + (1-n_i) \mathrm{ln} (1-n_i)\right], \eeq
where $n_i$ is given in Eq.~(\ref{occu}). With Eqs.~(\ref{energy})
and (\ref{entropy}), it is easy to obtain the nucleon contribution
to the free energy per nucleon $F$, \beq\label{fe:rmf} F = F_b/A =
(E_b\ -\ T S_b)/A. \eeq

\subsection{Uniform nuclear matter}
\label{subsec.uniform}

To make the paper self-contained, we give the formulas for uniform
matter in RMF model.  As we show below, at high temperatures or high densities the matter is uniform.  We include anti-nucleon terms which make a small contribution at very high temperatures.

The energy density of uniform nuclear matter is,
\beqn\label{energy_u} \varepsilon\ &=&\ \sum_{i=N, P}
\varepsilon^i_{kin}\ +\ \frac 1 2 [m_\sigma^2\sigma^2\ +\
m_\omega^2\omega_0^2\ +\ m_\rho^2\rho_{0,3}^2]\ \cr && +\ \frac 1
3 g_2\sigma^3\ +\ \frac 1 4 g_3 \sigma^4\ +\ \frac 3 4
c_3\omega_0^4, \eeqn where \beq \varepsilon^i_{kin}\ =\
\frac{2}{(2\pi)^3}\int d^3k E^*(k)[n_k(T)+\bar{n}_k(T)], \eeq with
effective mass $m^* = m + \Gamma_\sigma \sigma$,
and $E^*(k)=\sqrt{k^2+m^{*2}}$.  The occupation probabilities for particles $ n_k(T)$ and antiparticles $\bar{n}_k(T)$ are,
\begin{widetext} \beqn\label{occupation} n_k(T) &=& \frac {1}
{\mathrm{exp}{(E^*(k)+g_\omega\omega_0+g_\rho\tau^3\rho_{0,3}+\frac{\partial
\Gamma_\sigma}{\partial n}\rho_s \sigma-\mu)/T}+1},\\ \bar{n}_k(T)
&=& \frac {1}
{\mathrm{exp}{(E^*(k)-g_\omega\omega_0-g_\rho\tau^3\rho_{0,3}-\frac{\partial
\Gamma_\sigma}{\partial n}\rho_s \sigma+\mu)/T}+1}. \eeqn
\end{widetext}

The pressure of uniform nuclear matter is, \beqn\label{pressure_u}
P\ &=&\ \sum_{i=N,P} P_{kin}^i\ -\ \frac 1 2 m_\sigma^2\sigma^2 -
\frac 1 3 g_2\sigma^3 - \frac 1 4 g_3 \sigma^4\ \cr && +
\frac{\partial \Gamma_\sigma}{\partial n}\rho_s\sigma n +\ \frac 1
2m_\omega^2\omega_0^2\ +\ \frac 1 4 c_3\omega_0^4\ +\ \frac 1 2
m_\rho^2\rho_{0,3}^2, \cr && \eeqn where \beq P_{kin}^i\ =\
\frac{2}{3(2\pi)^3}\int d^3k
\frac{k^2}{\sqrt{k^2+m^{*2}}}[n_k(T)+\bar{n}_k(T)]. \eeq

The entropy density of uniform nuclear matter is,
\begin{widetext}
\beqn\label{entropy_u} s = -\frac{2k_B}{(2\pi)^3}\int
d^3k[n_k(T)\mathrm{ln}n_k(T)\ +\ (1-n_k(T))\mathrm{ln}(1-n_k(T))\
+\ \bar{n}_k(T)\mathrm{ln}\bar{n}_k(T)\ +\
(1-\bar{n}_k(T))\mathrm{ln}(1-\bar{n}_k(T))]. \eeqn
\end{widetext}

Using Eqs.~(\ref{energy_u}) and (\ref{entropy_u}) one can obtain the
free energy density per nucleon for uniform matter,
\beq\label{fe:u} F\ =\ (\varepsilon - Ts)/n_B. \eeq

\section{\label{numerics}Parameter set with density dependent coupling}

In this work we use the NL3 effective interaction \cite{NL3} that has been successful in reproducing ground state properties of stable nuclei and the saturation properties of
symmetric nuclear matter. The values of parameters in the NL3
effective interaction are listed in Table \ref{tab:para}.

\begin{table}[h]
\centering \caption{NL3 effective interaction. The nucleon masses
are $M$ = 939~MeV for both protons and neutrons and $c_3=0$ in
Eq.(\ref{lagrangian}). } \label{tab:para}\btab{cccccccc} \hline
\hline $\Gamma^0_\sigma$ & $g_\omega$ & $g_\rho$ & $g_2$ & $g_{3}$
& $m_\sigma$ &
$m_\omega$ & $m_\rho$\\
 & & & (fm$^{-1}$)  & & (MeV) & (MeV) & (MeV) \\
\hline
10.217&12.868&4.474&-10.431&-28.885& 508.194&782.5& 763\\
\hline \etab
\end{table}

As is well known, the mean field approach for pure neutron matter
is problematic at low densities because long range correlations are
important.  Neutron matter at low density is very close to a unitary
gas \cite{HS05b}, since the scattering length is
much larger than the inter-particle spacing, which is also
larger than the effective range of nuclear interaction.  To describe neutron
matter phenomenologically in the RMF framework, without losing its
success for the properties of nuclear matter, we introduce a
density-dependent scalar meson-nucleon coupling,
\beq\label{DD} \Gamma_\sigma\ =\ \left\{
\begin{array}{cc}
  \Gamma^0_\sigma, & n > n_0 \\
  \frac{\Gamma^0_\sigma}{1+\alpha} \left[(\frac{n+n_0}{2n})^{\frac 1 6} + \alpha\right],& n\leq n_0. \\
\end{array}\right. \eeq The two free parameters $n_0$ and $\alpha$
are determined by matching the energy of neutron matter to that of
a unitary gas at zero temperature $E_U$ \cite{unitarygas},
\beq\label{unitary} E_U\ =\ \xi\cdot\frac 3 5
\frac{k_F^2}{2m}\simeq 0.44\cdot\frac 3 5
\frac{k_F^2}{2m}, \eeq  where $k_F$ is the neutron Fermi momentum.  The best fitted values are $n_0=5\times 10^{-3}$ fm$^{-3}$ and $\alpha=1.2$.

\begin{figure}[htbp]
 \centering
 \includegraphics[height=8.5cm,angle=-90]{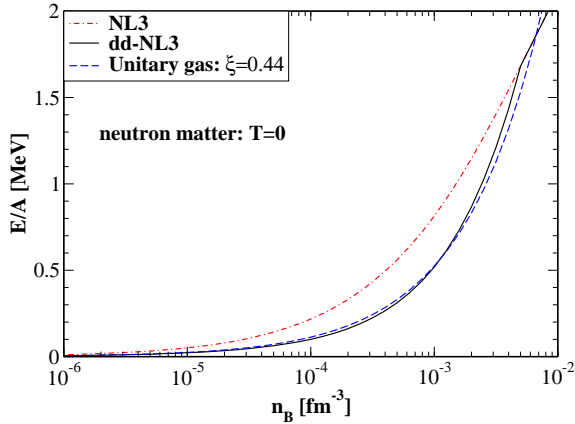}
\caption{(Color on line) Energy of pure neutron matter at $T$ = 0. The red curve
is from the original NL3 set. The black curve is for NL3 with a density
dependent $\sigma$-N coupling $\Gamma_\sigma$ as in
Eq.~(\ref{DD}).  The blue dashed line is the energy of a unitary gas, see
Eq.~(\ref{unitary}).}\label{fig:ldnm}
\end{figure}

In Fig.~\ref{fig:ldnm}, the energy of pure neutron matter at $T$ = 0
is shown for the original NL3 set, the modified NL3 set with a density
dependent $\sigma$-N coupling as in Eq.~(\ref{DD}), and the unitary
gas calculated by Eq.~(\ref{unitary}). The unitary gas gives lower
energy than the original NL3 result by about 0.2 MeV per particle,
due to the strong $S$ wave attractive interactions. This energy
difference is very relevant for matching a Virial expansion to the mean
field calculations, since the Virial expansion includes the long range
two-body neutron-neutron attractive interaction \cite{SHT10b,HS05} while the
normal mean field calculation does not. In the density range shown in the
figure, NL3 also gives a lower energy than the TM1 or FSUGold \cite{Fsugold} RMF 
parameter sets.  However, the density dependent NL3 can
fit the unitary gas result by tuning the coupling strength in the
attractive scalar meson channel.  Therefore, the density
dependent NL3 set describes successfully the properties of both
neutron rich matter and nuclear matter.  In the following when we
refer to NL3 set, we mean the density dependent NL3 set unless
otherwise specified.

\section{Computational Methodology}
\label{methodology}

In this section we describe our strategy for evaluating 
the equation of state.  We calculate the equation of state for the following partitioning of $T$, $n_B$, and $Y_P$ parameter space, see Table \ref{tab:phasespace}:
\begin{itemize}
\item{We use a step of 0.2 in $\log_{10}( T/ \ls {\rm{MeV}} \rs )$ for $\log_{10}( T/ \ls {\rm{MeV}} \rs )$ from -0.8 to 0, 
a step of 0.1 for $\log_{10}( T/ \ls {\rm{MeV}} \rs )$ from 0 to 1.1 and a step of 0.05 for $\log_{10}( T/ \ls {\rm{MeV}} \rs )$ above 1.1.  We have a total of 32 points for $T$ from 0.16 to 80 MeV. }
\item{ For temperatures $T$ below 15.8 MeV (where matter can be nonuniform) we use a step of 0.1 in $\log_{10}(n_B/ \ls {\rm{fm}}^{-3}\rs )$ for $\log_{10}(n_B/ \ls {\rm{fm}}^{-3}\rs )$ from -4.0 to 0.2.  We have a total of 43 points for $n_B$ from $ 10^{-4}$ to 1.6 fm$^{-3}$. }
\item{ For temperatures $T$ above 15.8 MeV (where matter is uniform) we use a step of 0.1 in $\log_{10}(n_B/ \ls {\rm{fm}}^{-3}\rs )$ for $\log_{10}(n_B/ \ls {\rm{fm}}^{-3}\rs )$ from -8.0 to 0.2.  We have a total of 83 points for $n_B$ from $ 10^{-8}$ to 1.6 fm$^{-3}$. }
\item{We use a step of 0.01 in proton fraction $Y_P$ for $Y_P$ from 0.05 to 0.56.  We aslo include $Y_P=0.0$.  This gives a total of 53 points for $Y_P$ from 0.0 to 0.56. } 
\end{itemize}
This partitioning gives a total of 40,248 points in the nonuniform Hartree region.  However for matter at higher temperatures, but still $T<15.8$ MeV, and/or lower proton fractions, Hartree results give higher free energies than corresponding results for uniform matter.  By roughly estimating the phase boundary, and keeping enough points to cross the transition density, see Section \ref{result} , we calculate the free energy for a reduced number of points in the nonuniform Hartree region, see Sec. \ref{subsec.nonuniform}, which includes a total of 17,021 points.  We also calculate free energies for uniform nuclear matter at a total of 90,478 points, see Sec. \ref{subsec.uniform}.

The most time is 
spent evaluating (temperature $T$, proton 
fraction $Y_P$, density $n_B$) points in the non-uniform Hartree mean 
field region. For each point we 
need to minimize the free energy of the W-S cell with respect to the cell 
radius which typically requires evaluation at 40 to 100 cell radii.  This 
minimization can be complicated by the existance of local minima.  
For each cell  size, we need to solve the mean fields self consistently.
We have  already developed the code for this minimization 
(see Ref. ~\cite{HS08}) which is slightly modified in this work to accomodate 
the density dependent coupling in RMF. 

The mean fields provide potentials for the individual nucleons in the W-S 
cell which obey the Dirac equation. The Dirac equation is solved by 
a fourth order Runge-Kutta method with shooting techniques. For nuclear matter 
at finite temperature, there could be hundreds of nucleons that populate 
thousands of levels according to Fermi-Dirac statistics. For each level, 
we need to solve the Dirac equation. The potentials for the nucleons in 
the Dirac Eq.~(\ref{Dirac}) are various meson mean fields which obey the 
extended Klein-Gordon (K-G) equation. The source terms for the K-G 
equations are provided by various nucleon density terms in Eq.~(\ref{source}).
Given the nucleon density terms, the K-G equations are solved by a 
Green's function method, which updates the meson mean fields. The updated 
mean fields can now be used 
to solve the nuclear levels and nucleon densities again.  This process is 
repeated until full self-consistency is achieved in both the mean fields 
and the nuclear levels.

Computationally, the problem to be solved is embarrassingly parallel 
because each point of density, temperature and 
proton fraction is independent of the others.
A total number of $\sim$ 17,000 
independent tasks must be run, where each task
calculates the required quantities at a single point in the phase 
space. 
Unfortunately, the run time on an individual task varies from a few 
minutes to more than 24 hours, depending on the number of iterated 
cell radii, and the number of nucleon energy levels included.  

Each point in the phase space was mapped to a unique integer that we 
refer to as the job index.  A file, {\it runlist}, was prepared with a 
list of job indicies for the whole phase space, and a single character 
(A=available, R=running, r=Re-running,
C=complete, T=time-limited and F=failed) that gives the status of 
calculations for that job index.  An MPI parallel wrapper code manages 
the running of the many requested  tasks. 
Typically, one parallel job requests a set of compute 
cores (usually 256). Each MPI rank,
using a single CPU core, is assigned one job index 
corresponding to one point in the phase space and it evaluates 
the required quantities.

Initially, rank zero of the MPI job 
\begin{itemize}
\item{locks the job listing file {\it runlist}}
\item{reads {\it runlist} until a list of available 
tasks is filled}
\item{closes {\it runlist} and releases the lock}
\item{passes a job index to each MPI rank and begins the calculation
for that job index.}
\end{itemize}

When the calculation completes (or time-limits or fails) for a given 
MPI rank, the status character for the job index in 
{\it runlist} is modified appropriately. The now available 
MPI rank will search {\it runlist} for next available task and the 
calculation restarts for the new job index. 
Since completion occurs asynchronously file locking is not used for this
part of the process.

A simple batch job runs through the points in phase space. A wall clock 
limit (48 hours) larger than the average run time is used. Each rank of 
the MPI job can run a series
of points via above procedure, efficiently using each available core for 
the requested wall clock period. 
One job per core is running when the wall clock 
limit is reached. These jobs are
identified by being left in the "R" state after the batch job completes.
Using this scheme we have achieved 85\%
efficiency in CPU usage. Specifically, 85{\%} of all jobs ended in the
"C" state rather than the "T" or "R" state. After the {\it runlist} has 
been searched once, the remaining jobs have "R" or "T" state. Then these 
remaining jobs are resubmitted via the MPI wrapper code, requesting longer 
time limit (typically 7 days) but fewer CPU cores. This procedure allows 
us to calculate $>$ 99 $\%$ of the points in the {\it runlist} file.

\section{\label{result}Results}

In this section we discuss our results for various regions of parameter
space.  First, the uniform matter EOS at zero temperature is
presented.  Second, we discuss the free energy per nucleon for mean field calculations of nonuniform matter.  Finally, we show the density distributions of neutrons and protons inside W-S cells.

\subsection{Uniform matter at zero temperature}

\begin{figure}[htbp]
 \centering
 \includegraphics[height=8.5cm,angle=-90]{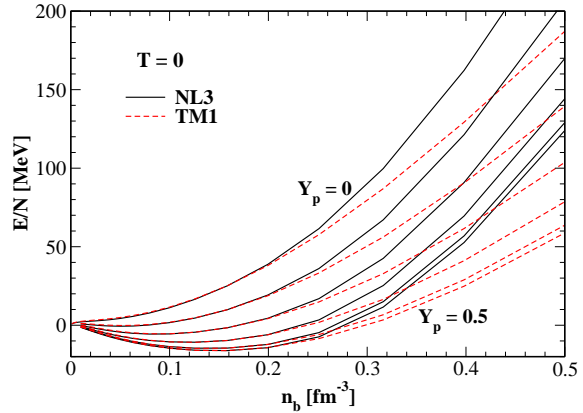}
\caption{(Color on line) Energy of uniform matter at zero temperature with
different proton fractions, $Y_P$ = 0, 0.1, 0.2, 0.3, 0.4 and
0.5.}\label{fig:zeroT}
\end{figure}

Figure~\ref{fig:zeroT} shows the equation of state of uniform matter
at zero temperature with different proton fractions $Y_P$ = 0, 0.1, 0.2, 0.3, 0.4, and 0.5. The
solid curves are for the NL3 parameter set, which is used in our RMF
calculations.  The dashed curves show results for the TM1 interaction, which
is used in the equation of state obtained by H. Shen et al
\cite{Shen98}.  The two sets agree to a great extent for densities
below 0.2$\sim$0.25 fm$^{-3}$, depending the value of $Y_P$. Above
these densities, NL3 gives a much stiffer equation of state for
uniform matter. This serves as one motivation for our choice of
the NL3 parameters: to explore the equation of state with a stiffer symmetry energy.

\subsection{Free energy and phase boundaries}

In Fig.~\ref{fig:fe}, the free energy per nucleon $F/A$ is shown as a function of density $n_B$ at $T$ = 1, 3.16, 6.31 and 10 MeV. At intermediate densities, $F/A$ is calculated from Eq.~(\ref{fe:rmf}) for W-S cells using Hartree mean field calculations. At high densities, $F/A$ is calculated from Eq.~(\ref{fe:u}) for uniform matter. 
The transition (as the density grows) is found at the density where
uniform matter gives a lower free energy.
In each panel, the (red) solid curves give
the transition densities to uniform matter. 
The transition densities increase as proton fraction grows. Non-uniform matter can exist until higher densities in more symmetric nuclear matter.
At density around 0.16 fm$^{-3}$, there is always a minimum in
the free energy per nucleon, as long as the proton fraction is not too small and
temperature not too high. This is the manifestation of saturation density in nuclear matter.

\subsection{Density distributions inside Wigner-Seitz cells}


The Hartree mean field calculation provides detailed
wavefunctions for nucleons in the non-uniform phase. In
our W-S approximation at intermediate densities, we find a ``spherical pasta" phase where the proton density distribution forms a  shell state with a reduced density in the center.  This reduces the large coulomb repulsion between protons and was discussed in \cite{HS08}.
In this section we discuss density distributions inside the W-S cells,
both for normal nuclei and for these shell states.

In Fig.~\ref{fig:yp0.1}, neutron and proton distributions,
inside the W-S cell, are shown for four different baryon
densities, with $T$ = 1 MeV and $Y_P$ = 0.1.  At very low density
$n_B$ = 0.002 fm$^{-3}$, the Hartree calculation has a minimum for $Z=39$ protons and $A= 390$ nucleons.  Most of the neutrons are located within 10 fm of the cell center, although the
W-S cell radius is around 31 fm.  A small fraction of the neutrons extend to
the edge of cell since this is an extremely neutron rich system.  As the
density rises to 0.02 fm$^{-3}$, the W-S cell has $Z=42$, $A= 420$, and the
neutron density at large $r$ becomes much greater.  The W-S cell radius drops to
17.5 fm because the lattice becomes more closely packed as the density increases. At a density of 0.05 fm$^{-3}$, the W-S cell has $Z=413$ and $A= 4130$ and forms a shell state with both inside and outside surfaces. This has been
discussed in our early paper \cite{HS08}.  As a result, the W-S cell
radius become larger.  At the higher density of 0.063 fm$^{-3}$, the system becomes uniform.

In Fig.~\ref{fig:y3n02}, the distribution of neutrons and
protons are shown for $n_B$ = 0.020 fm$^{-3}$ and proton fraction $Y_P$ =0.3.  At low temperature $T$ = 1 MeV, where $Z=85$, $A=282$, the density distributions are similar to those for normal isolated nuclei.  As the temperature rises to 3.16 and 6.31 MeV, the size of W-S cell remains nearly fixed, but the neutron density increases with temperature at large radius.  This is due to excitation of states with high angular momentum and/or large main quantum number as the temperature rises.  When the temperature rises to 10 MeV, the proton density also rises at large $r$, accompanied by an increase of the W-S cell size, with $Z=123$, $A=410$. At even higher temperature, the nucleus melts and uniform matter appears.

Similar to Fig.~\ref{fig:y3n02} but at higher $n_B$ = 0.050 fm$^{-3}$ and proton fraction $Y_P$ = 0.45, the density distribution of neutrons and protons are shown for four different
temperatures in Fig.~\ref{fig:y45n05}.  Here a shell state exists up to high temperatures.  At low temperatures $T$ = 1, 3.16 MeV, $Z=1315$, $A = 2922$ and the shell state has inside and outside voids.  As the temperature rises, nucleons populate both the inside and outside voids, due to thermal excitations.   Finally, the size of the shell state shrinks at high temperature so that $Z=648$, $A=1440$ at $T$ =10 MeV.

\section{\label{summary}Summary and Outlook}

In this paper we present large scale relativistic mean field calculations for nuclear matter at intermediate and high densities. We use a density dependent modification of the NL3 interaction in a spherical Wigner-Seitz approximation.  Nuclear shell effects are included.  We calculate the free energy, and tabulate the resulting equation of state at over 107,000 grid points in the proton fraction range $Y_P$ = 0 to 0.56.  For low temperatures $T$ = 0.16 to 15.8 MeV we calculate for the density range $n_B$ = 10$^{-4}$ to 1.6 fm$^{-3}$.  For high temperatures $T$ = 15.8 to 80 MeV, where the matter is uniform, we calculate for the larger density range $n_B$ = 10$^{-8}$ to 1.6 fm$^{-3}$.  These calculations took over 6000 CPU days.

We solve for the nucleon Dirac wave functions and meson mean fields self-consistently.  
This allows us to study how the distribution of neutrons and protons inside a Wigner Seitz cell evolve with density and temperature.  We find a large variety of possible sizes and shapes for these distributions.

This paper provides part of our results for an equation of state which will cover a broad range of temperatures, densities, and proton fractions.  In the future, we plan to study low density nuclear matter using a Virial expansion for a non-ideal gas consisting of nucleons and thousands of species of nuclei. Then, we will generate a complete thermodynamically consistent equation of state by matching the low density and higher density results.  This equation of state avoids the Thomas Fermi and variational approximations of the H. Shen et al. equation of state and is exact in the low density limit. It can be used in supernova and neutron star merger simulations. Finally, in future work we will generate equations of state using other modern relativistic mean field interactions such as FSUGold \cite{Fsugold}.

\begin{widetext}

\begin{figure}[htbp]

 \centering
 
  \includegraphics[height=8.5cm,angle=-90]{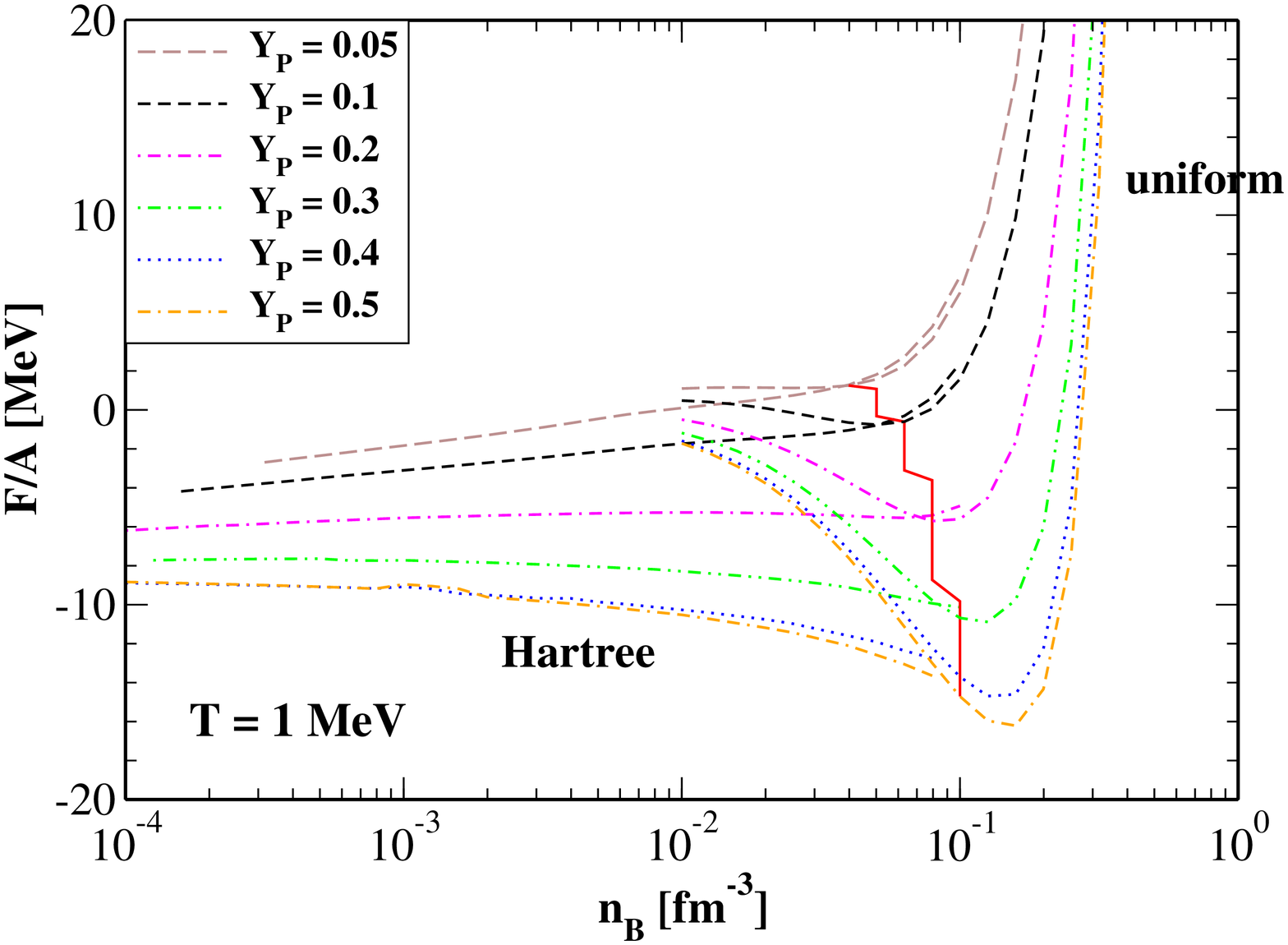}
 \includegraphics[height=8.5cm,angle=-90]{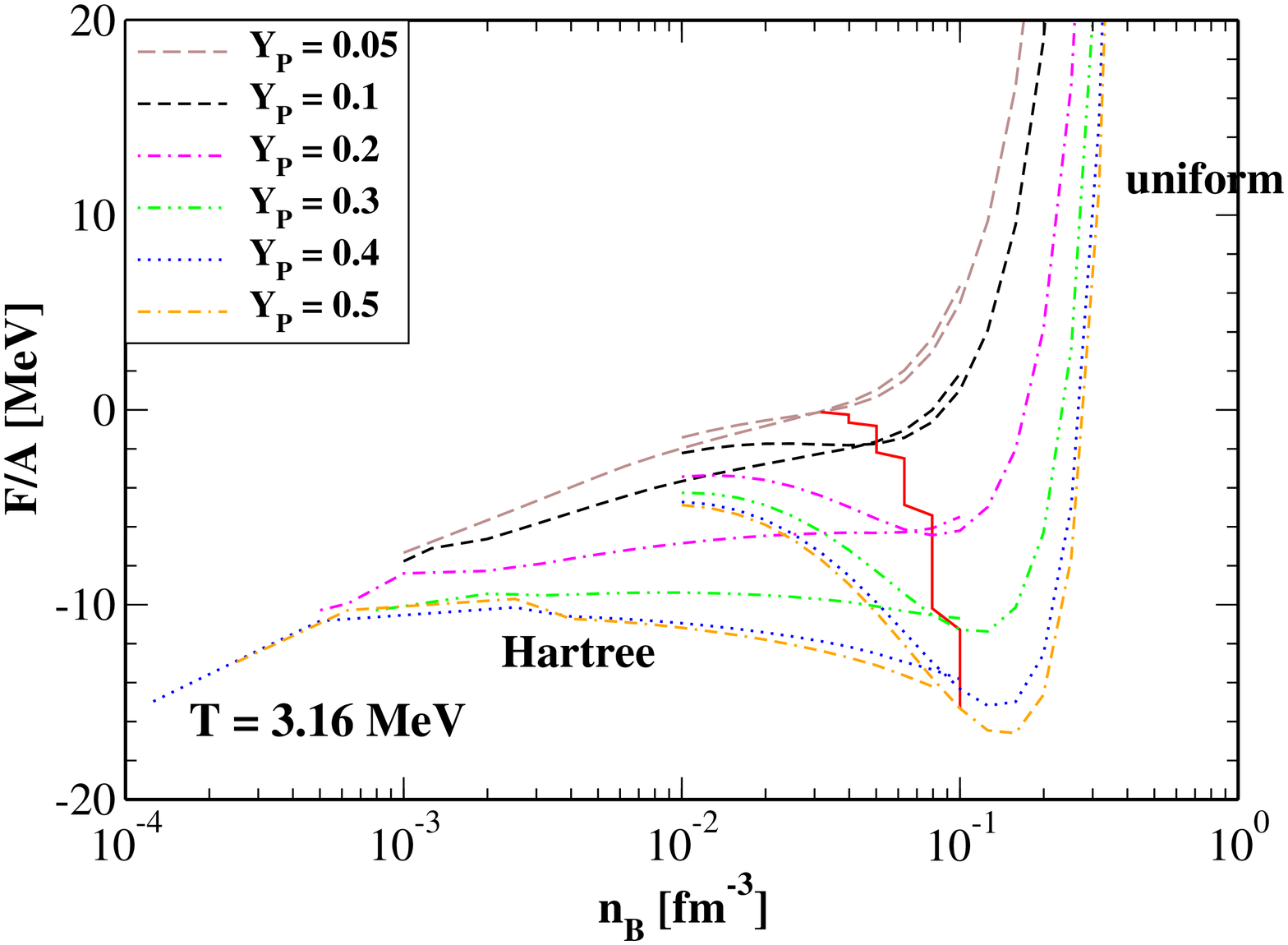}
 \includegraphics[height=8.5cm,angle=-90]{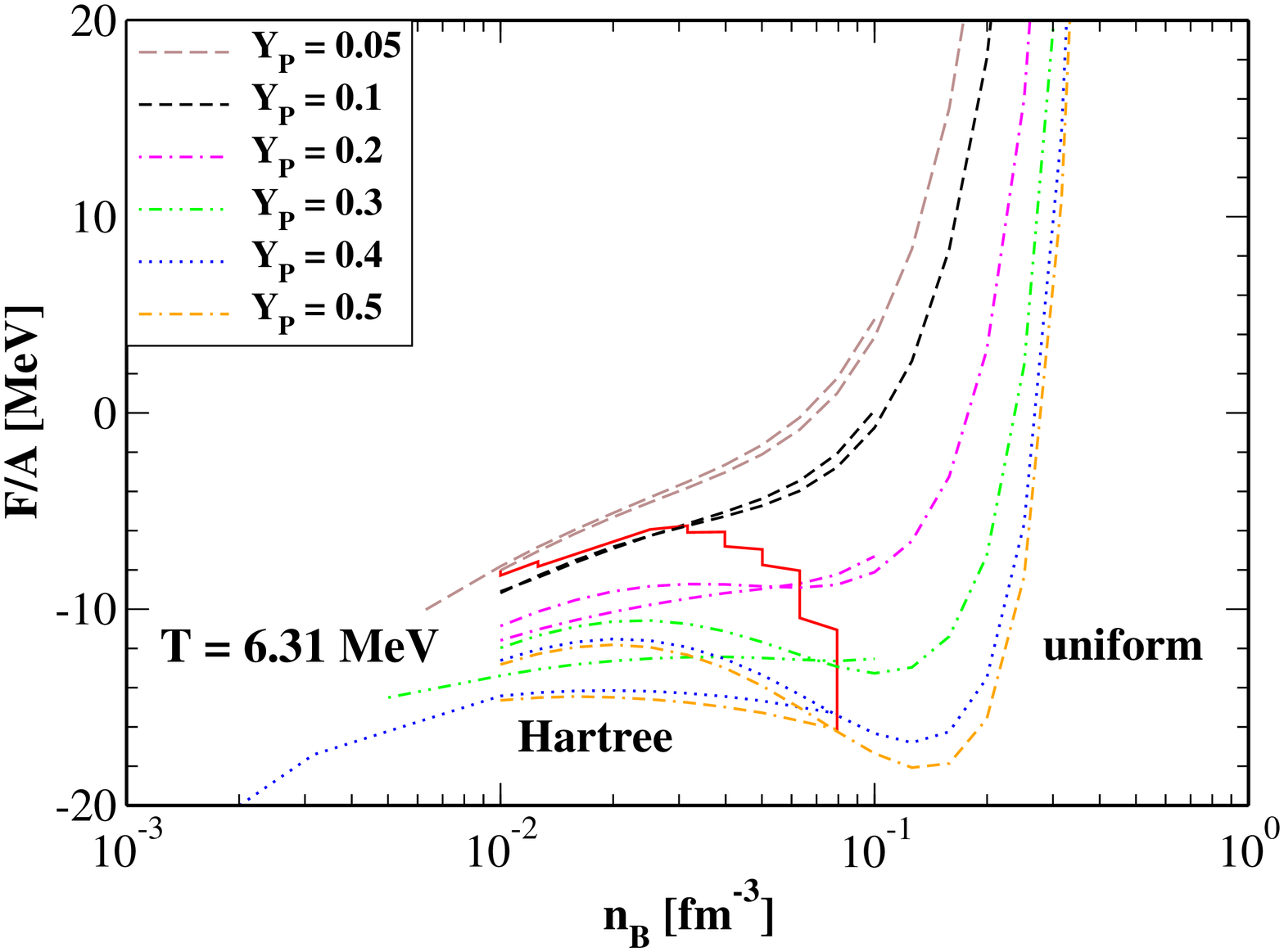}
 \includegraphics[height=8.5cm,angle=-90]{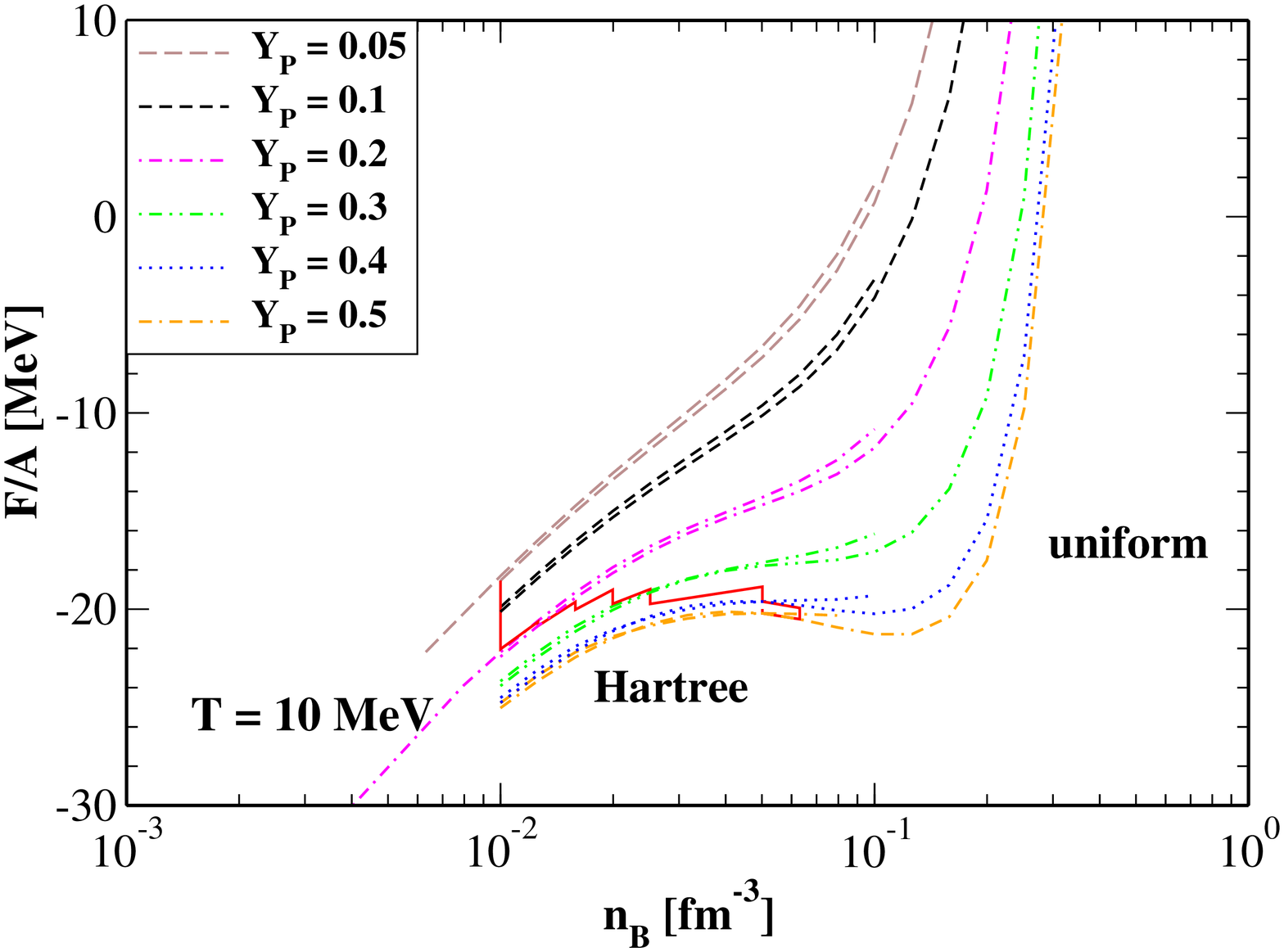}
\caption{(Color on line) Free energy per nucleon of nuclear matter at different
temperature and proton fractions. }\label{fig:fe}
\end{figure}



\begin{figure}[htbp]
 \centering
 \includegraphics[height=13cm,angle=-90]{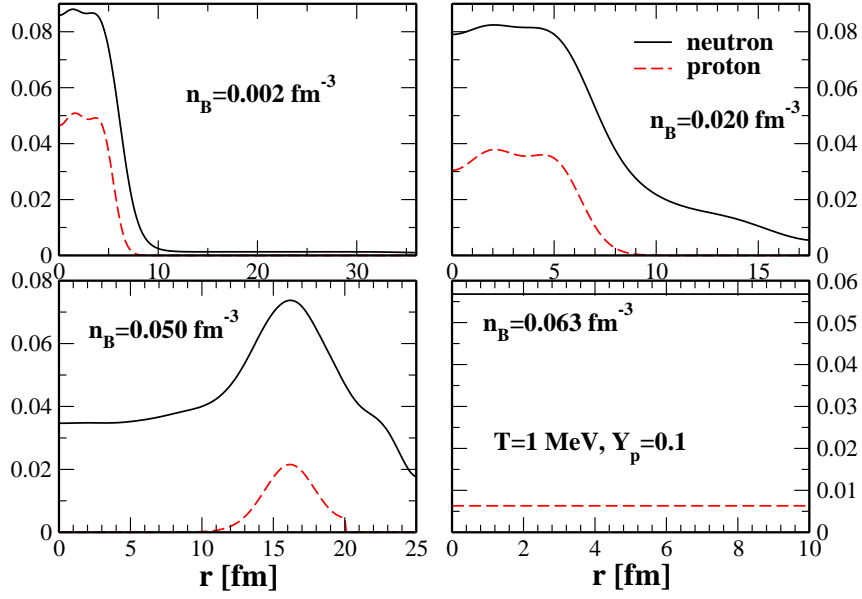}
\caption{(Color on line) Density distribution of neutrons, protons inside W-S cell
for four different baryon densities at $T = 1$ MeV and proton
fraction $Y_P$ = 0.1.}\label{fig:yp0.1}
\end{figure}
\begin{figure}[htbp]
 \centering
 \includegraphics[height=13cm,angle=-90]{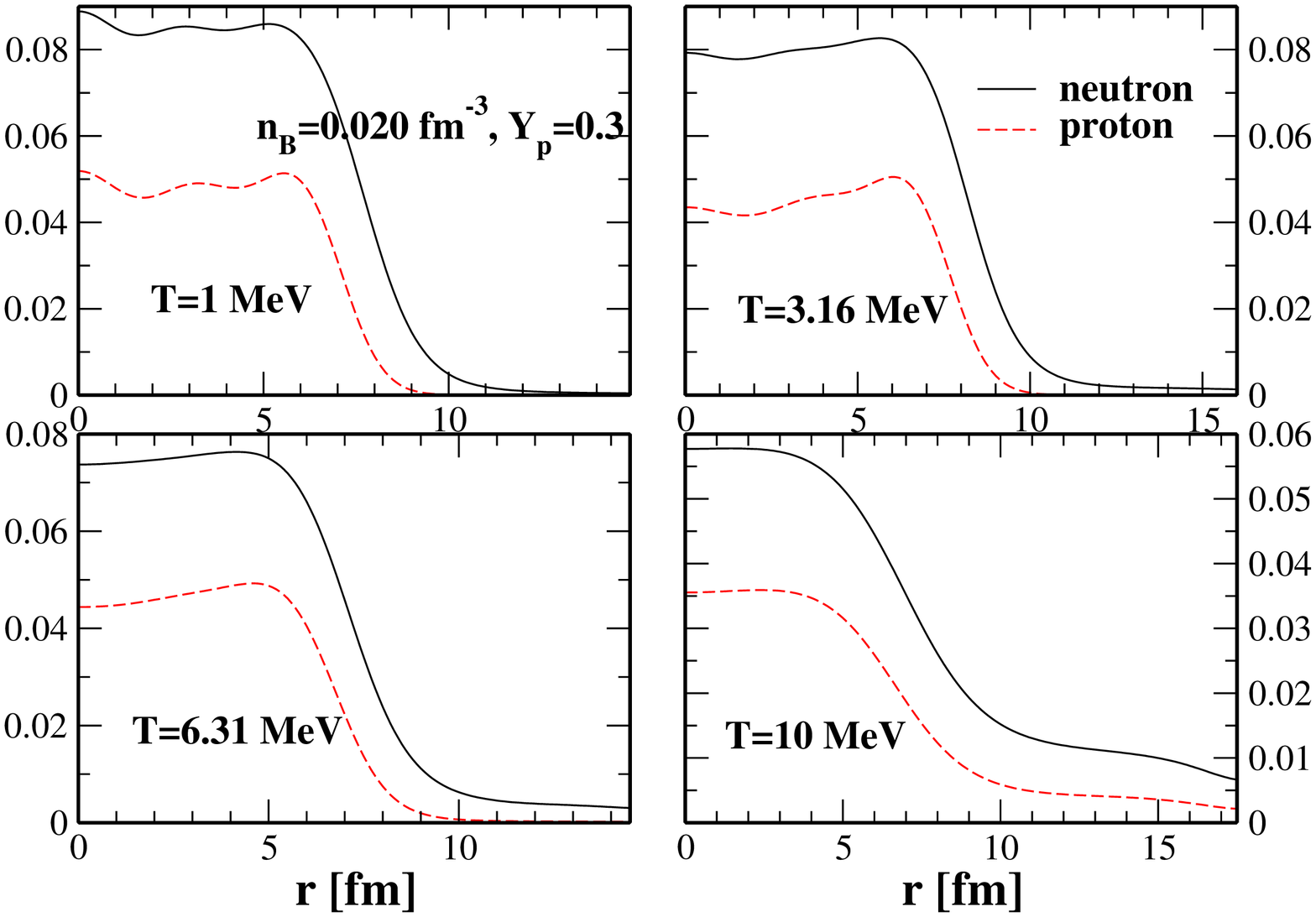}

\caption{(Color on line) Density distribution of neutrons, protons inside W-S cell
for four different temperatures with $n_B$ = 0.020 fm$^{-3}$ and
proton fraction $Y_P$ = 0.3.}\label{fig:y3n02}
\end{figure}



\begin{figure}[htbp]

\centering

 \includegraphics[height=13cm,angle=-90]{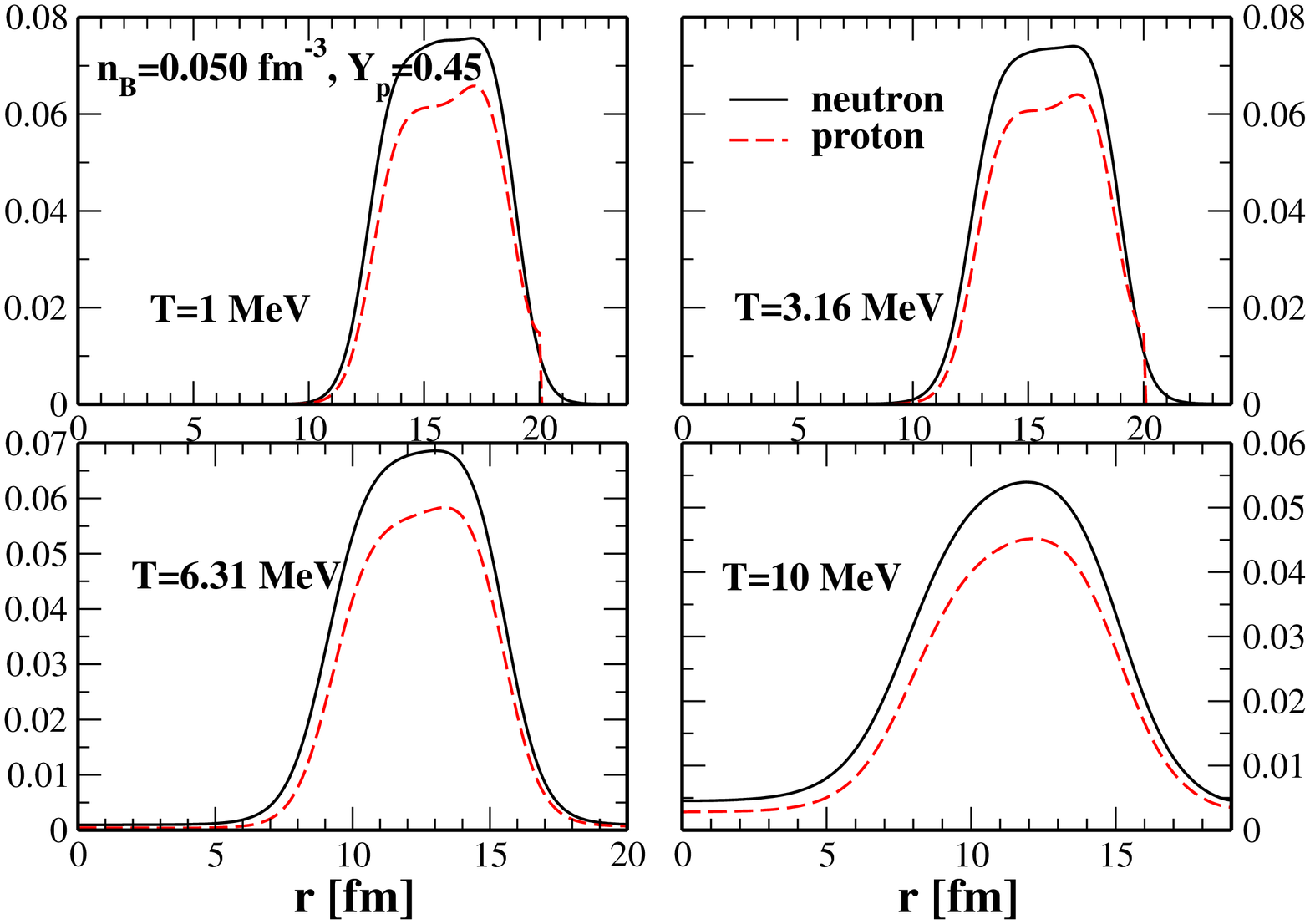}
\caption{(Color on line) Density distribution of neutrons, protons inside W-S cell
for four different temperatures at $n_B$ = 0.050 fm$^{-3}$ and
proton fraction $Y_P$ = 0.45.}\label{fig:y45n05}
\end{figure}

\end{widetext}

\section{Acknowledgement}

We thank Lorenz H$\ddot{\mathrm{u}}$edepohl, Thomas Janka, Jim Lattimer, Andreas Marek, Evan O'Connor, and Christian Ott for helpful discussions. This work was
supported in part by DOE grant DE-FG02-87ER40365. This material is based upon work supported by the National Science Foundation under Grants No. ACI-0338618l, OCI-0451237, OCI-0535258, OCI-0504075 and CNS-0521433.
This research was supported in part by Shared University Research grants from IBM, Inc. to Indiana University and by the Indiana METACyt Initiative.
The Indiana METACyt Initiative of Indiana University is supported in
part by Lilly Endowment.

\end{document}